\documentclass[epsf]{elsart}
\usepackage{graphicx}% Include figure files
\usepackage{dcolumn}% Align table columns on decimal point
\usepackage{bm}% bold math
\usepackage{amsmath}
\usepackage{amssymb}
\def\dis{\displaystyle}
\def\beq{\begin{equation}}
\def\eeq{\end{equation}}
\def\[{\left[}
\def\]{\right]}
\def\ni{\noindent}

\def\dis{\displaystyle}
\def\beq{\begin{equation}}
\def\eeq{\end{equation}}
\def\[{\left[}
\def\]{\right]}
\def\ni{\noindent}

\input epsf

\begin{document}

%\DeclareGraphicsRule{.eps.gz}{eps}{.eps.bb}{'gunzip -c #1}    

%\preprint{APS/123-QED}

\begin{frontmatter}

\title{Probing the charged Higgs quantum numbers through the decay 
$H^{+}_\alpha \rightarrow W^{+}h^{0}_\beta$}
\author[fcfmbuap]{J. L. Diaz--Cruz},
\author[ifunam]{Olga F\'elix--Beltr\'an},
\author[ciaii]{ J. Hern\' andez--S\' anchez},
\author[fcfmbuap]{E. Barradas--Guevara},
\address[fcfmbuap]{Academic Body for Particles
 and Fields, Facultad de Ciencias F\'{\i}sico Matem\'aticas, BUAP,
      Apartado Postal 1152, 72000 Puebla, Pue., M\'exico }
\address[ifunam]{Instituto de F\'{\i}sica, UNAM, Apdo. Postal 20-364,
M\'exico 01000 D.F., M\'exico}
\address[ciaii]{Centro de Investigaci\'on Avanzada en Ingenier\'{\i}a 
Industrial, Universidad Aut\'onoma del Estado de Hidalgo,
 Carretera Pachuca-Tulancingo Km. 4.5, Ciudad Universitaria, C.P. 42020
 Pachuca, Hgo., M\'exico}
 
\begin{abstract}
\noindent The vertex $H^{+}_\alpha W^{-}h^{0}_\beta$, involving the gauge 
boson $W^\pm$ and the  charged ($H^{\pm}_\alpha$) and neutral Higgs 
bosons ($h^{0}_\beta$), arises  within the context of many  extensions of 
the SM,
and it can be used to probe the quantum numbers of the Higgs multiplet.
After presenting a general discusion for the expected form of this 
vertex with arbitrary Higgs representations, we discuss its strength
for several specific models, which include: $i)$ the Two-Higgs Doublet
Model (THDM), both the generic and the SUSY case, and 
$ii)$ models with additional Higgs triplets, 
including  both SUSY and non-SUSY cases. We find  that in these models, 
there are regions of parameters where the decay  
$H^{+}_\alpha \to  W^{+}h^{0}_\beta$, 
is kinematically allowed, and reaches Branching Ratios (BR) that may be 
detectable, thus 
allowing to test the properties of the Higgs sector. 
\end{abstract}

\end{frontmatter}

\maketitle
%%%%%%%%%%%%%%%%%%%%%%%%%%%%%%%%%%%%%%%%%%%%%%%%%%%%%%%%%%%%%%%%%%%%%%%%%
\section{Introduction}
%%%%%%%%%%%%%%%%%%%%%%%%%%%%%%%%%%%%%%%%%%%%%%%%%%%%%%%%%%%%%%%%%%%%

The Higgs spectrum of many well motivated extensions of the Standard Model 
(SM) often includes a charged Higgs, whose detection  at future colliders 
would constitute a clear evidence of a Higgs sector beyond the 
minimal SM \cite{kanehunt}. 
 In particular, the two-Higgs doublet model (THDM) has been extensively 
studied as a prototype of a Higgs sector that includes a 
charged Higgs boson ($H^\pm$)\cite{kanehunt}. 
However, a definitive test of the mechanism of electroweak symmetry  
breaking will require further studies of the complete Higgs spectrum.  
In particular, probing the properties of charged Higgs could help to 
find out whether it is indeed associated with a weakly-interacting theory, 
as in the case of the popular minimal SUSY extension of the SM (MSSM) 
\cite{susyhix}, or to a strongly-interacting scenario \cite{stronghix}. 
Furthermore, these tests should also allow to probe the symmetries of 
the Higgs potential, and to determine whether the charged Higgs belongs 
to a weak-doublet or to some larger multiplet.

Decays of a charged Higgs boson have been studied in the literature, 
including the radiative modes $W^+\gamma, W^+Z$ \cite{hcdecay}, 
mostly within the context of the THDM,  or its MSSM incarnation, 
and more recently for the effective Lagrangian extension of the THDM 
\cite{ourpaper}. Charged Higgs production at hadron colliders was studied 
long ago \cite{ldcysampay}, and recently more systematic calculations of 
production processes at  LHC have been presented \cite{newhcprod}. 
Current bounds on charged Higgs mass can be obtained at Tevatron, 
by studying the top decay $t \to bH^+$, which already eliminates 
some region of parameter space \cite{LHCbounds}, whereas LEP 
bounds give approximately $m_{H^+} > 100$ GeV \cite{LEPbounds}.

On the other hand, the vertex $H^+_\alpha W^- h^0_\beta$, 
deserves special attention 
because it can give valuable information about the underlying structure of 
the gauge and scalar sectors. In the first place,  the decay mode 
$H^+_{\alpha}\to W^+ h^0_{\beta}$ could be detected 
at the future Large Hadron Collider (LHC) as it was 
claimed in reference \cite{hcwhdetect} within the context of the MSSM. Furthermore, 
the vertex $H^+_\alpha W^- h^0_\beta$, can also induce 
the associated production of $H^+_\alpha + h^0_\beta$ 
at hadron colliders, through a 
virtual $W^{+*}$ in the s-channel, which could become a relevant 
production mechanism for heavy charged Higgs bosons.

In this paper we are interested in studying thoroughly the physics behind 
this important vertex. Its organization goes as follows: in section 2, 
we present a general analysis of the case when the Higgs sector
includes arbitrary Higgs representations; we derive a general formula 
for the vertex $H^+_\alpha W^- h^0_\beta$ in terms of the isospin components  
and the hypercharge of the Higgs multiplet. Then, in section 3, we apply this 
general discussion to study the charged Higgs vertex within 
the THDM and  the minimal SUSY extension of the SM (MSSM).
Results for the BR of the charged Higgs decay
in the MSSM are presented including the leading radiative corrections. 
In section 4, we discuss the strength of the vertex for an extended 
supersymmetric model that includes a complex Higgs triplet; we perform
a numerical analysis to search for values of Higgs masses 
that allow the decays $H^+_\alpha \to W^+ h^0_\beta$ to procced. 
Finally, in section 5  we present  our conclusions. 

%%%%%%%%%%%%%%%%%%%%%%%%%%%%%%%%%%%%%%%%%%%%%%%%%%%%%%%%%%%%%%%%%%%%%%%
\section{General Analysis} 
%%%%%%%%%%%%%%%%%%%%%%%%%%%%%%%%%%%%%%%%%%%%%%%%%%%%%%%%%%%%%%%%%%%%%%%

Let us consider a Higgs sector that includes several Higgs multiplets 
$\Phi_\alpha$, which transforms under $SU(2) \times U(1)$ with isospin 
($T_\alpha$)  and  hypercharge $(Y_\alpha)$. The components of the isospin 
$T_\alpha$, $T_{i \alpha }$ ($i=1,2,3$) are $n$-dimensional representations 
of $SU(2)$, and satisfy the algebra
 $[T_{i \alpha }, T_{j \alpha }]  = i \epsilon_{ijk} T_{k \alpha }$.
 From the operators $T_{i \alpha}$  one can
define the raising and the lowering operators 
$T^\pm_\alpha = T_{1 \alpha} \pm i T_{2\alpha }$,
which will be used next.

 The kinetic terms of the Higgs multiplets are given by  
\begin{equation}
{\mathcal L}_{K}=
\sum_\alpha (D^{\mu} \Phi_{\alpha})^{\dag } (D_{\mu} \Phi_{\alpha}), \label{lk}
\end{equation}
\ni where $D_\mu$ denotes the covariant derivative, and it takes the following 
form for a general multiplet, 
\begin{equation}
\begin{array}{rcl}
D_\mu & = & \partial_\mu -\displaystyle{\frac{ig}{\sqrt{2}}}(T^+ W^+_\mu + T^- W^-_\mu )
 - \displaystyle{\frac{ig}{c_w}}(T_3-s_w^2 Q)Z_\mu-ieQ A_\mu , \label{Dmu} 
\end{array}
\end{equation}
\ni where  $Q$ is the charge operator and the hypercharge $Y$ is 
normalized to satisfy the relation $Q= T_3+Y/2$. Equation (1)
will induce the gauge boson masses and the Higgs-gauge vertices
after spontaneous symmetry breaking (SSB).

%%%%%%%%%%%%%%%%%%%%%%%%%%%%%%%%%%%%%%%%%%%%%%%%%%%%%%%%%%%%%%%%%%%%%%%%%
\subsection{The Goldstone bosons}
%%%%%%%%%%%%%%%%%%%%%%%%%%%%%%%%%%%%%%%%%%%%%%%%%%%%%%%%%%%%%%%%%%%%%%%%%%

The Higgs multiplet $\Phi_\alpha$  can be expanded 
in terms of the spinors ${\chi}^{n(+)}$, eigenstates
of $T^2_\alpha$ and $T_{3\alpha}$, as follows:
\begin{equation}
\begin{array}{rcl}
\Phi_\alpha & = &
 \sum_{n} 
\phi_{\alpha}^{n(+) } \chi_\alpha^{n(+) } , \qquad  
(\Phi_\alpha )^\dag  = 
 \sum_{n} 
(\phi_{\alpha}^{n(+) })^* (\chi_\alpha^{n(+) })^\dag ,   \label{mul} 
\end{array}
\end{equation}
\ni where the components $\phi_{\alpha}^{n(+) } $ denote the scalar state 
with $n$ units of the electric charge, and include: 
$\phi_{\alpha}^{0 }, \phi_{\alpha}^{\pm }, \phi_{\alpha}^{\pm \pm },... $,
for $n=0, \pm 1, \pm 2,...$. The spinors $\chi_\alpha^{n(+)} $, being  eigenstates of $T^2_\alpha$ and 
$T_{3 \alpha}$, satisfy the following relations, 
\begin{equation}
\begin{array}{rcl}
(\chi_\alpha^{m(+)})^\dag (\chi_\beta^{n(+)}) & = & \delta_{\alpha, \beta}
\delta^{m,n},   \label{normalization}  \\
T_{3 \alpha} \chi_\alpha^{m(+)} & = &T_{3 \alpha}^m  \chi_\alpha^{m(+)}, 
\qquad \qquad T_{3 \alpha}^m = T_{\alpha}, T_{\alpha}-1 , ..., -T_{\alpha}
 \\
T_{\alpha}^\pm \chi_\alpha^{m(+)} & = & 
T_{ \alpha}^{\pm, m} \chi_\alpha^{m\pm 1(+)}, \qquad
T_{ \alpha}^{\pm, m}  =  \bigg[ (T_\alpha \mp T_{3 \alpha }^m)
( T_\alpha \pm T_{3 \alpha }^m +1 )\bigg]^{1/2},  
\end{array}
\end{equation}
\ni where $ T_{3 \alpha}^m $ and  $ T_{ \alpha}^{\pm, m} $ are the
eigenvalues of the operators $ T_{3 \alpha} $ and  $ T_{ \alpha}^{\pm} $ respectively.

After SSB, the neutral Higgs components acquire vacuum expectation 
values (v.e.v.'s), and one can write:
\begin{equation}
\begin{array}{rcl}
<\Phi_\alpha> & = & 
v_{\alpha} \chi_\alpha^{0 } , \qquad
<\Phi_\alpha >^\dag  = 
v_{\alpha}^* (\chi_\alpha^{0 })^\dag .   \label{vev}
\end{array}
\end{equation}
\ni While in some particular model it is possible that some of 
the $<\Phi_\alpha> $  could be absent, in the following we shall assume 
that $v_\alpha^* = v_\alpha$, which corresponds to a CP-invariant 
vacuum. Thus, in order to obtain the Higgs mass matrices and interactions
we need to make the substitution 
$\Phi_\alpha \to  \Phi_\alpha + <\Phi_\alpha> $ into the
lagrangian (1). 

Expanding equation (\ref{lk}), gives the following 
expression for the linear term involving the charged gauge boson,
\begin{equation}
\begin{array}{rcl}
({\mathcal L}_K)_{W^\mp \phi^\pm \phi^0 } & = &
\displaystyle{\frac{ig}{\sqrt{2}}} W^-_\mu \sum_\alpha \bigg[ \Phi_\alpha^\dag 
T^- (\partial_\mu \Phi_\alpha ) 
-  (\partial_\mu \Phi_\alpha )^\dag T^-  
\Phi_\alpha  \bigg]   \\
& - & \displaystyle{\frac{ig}{\sqrt{2}}} W^+_\mu \sum_\alpha \bigg[  
  (\partial_\mu \Phi_\alpha )^\dag T^+  
\Phi_\alpha  -  \Phi_\alpha^\dag 
T^+ (\partial_\mu \Phi_\alpha )  \bigg] . \label{gWHh}         
\end{array}
\end{equation}
\ni From this equation one can identify the combination of fields 
that correspond to the charged Goldstone boson $G_W^\pm $, by
separating terms  of the form  
$i m_W (W_\mu^- \partial^\mu G_W^+ - W_\mu^+ \partial^\mu G_W^- )$, 
which leads to
\begin{equation}
\begin{array}{rcl}
G_W^+ = \displaystyle{\frac{g}{\sqrt{2} m_W}} \sum_\alpha 
\bigg[ (T^+ <\Phi_\alpha>)^\dag \Phi_\alpha 
-\Phi_\alpha^\dag  T^- <\Phi_\alpha> \bigg]  .
\end{array}  
\end{equation}
The expression for the charged Goldstone bosons can be written then in terms 
of the components $\phi_\alpha^\pm$, as follows:
\begin{equation}
G_W^+ = \frac{g}{\sqrt{2} m_W} \sum_\alpha v_\alpha
\bigg[    T^{+,0}_\alpha  \phi_\alpha^+ 
-  T^{-,0}_\alpha (\phi_\alpha^-)^* \bigg] C_{ Y_\alpha } .
\end{equation}
For the cases when $T_\alpha$ is integer and  $Y=0$
(i.e. in real representation  ), one has:  
$ T^{+,0}_\alpha =  T^{-,0}_\alpha= \sqrt{T_\alpha(T_\alpha+1)}$.  
On the other hand, when  $2T_\alpha = Y_\alpha$, 
which corresponds to a complex representation,  
and $T_\alpha$ could be either integer or half-integer 
(e.g., Higgs doublets with $Y=1$ or triplets with $Y=2$),
one has $T^{+,0}_\alpha = 0$  for $Y_\alpha < 0$  and  $T^{-,0}_\alpha= 0$
when $Y_\alpha > 0$.  Furthermore, one can fix  
the following phase convention:   
 $\phi_{\alpha, T}^+ = - (\phi_{\alpha ,T}^-)^*$. We also notice that when $T_\alpha$ is integer
 and $Y_\alpha=0$, $<\Phi_{\alpha} (T, Y)>$  can contribute to $m_W$. 

Thus, the most general expression for the charged Goldstone bosons
for a Higgs sector that includes an arbitrary number of multiplets 
$\Phi_{\alpha} (T, Y) $ (either with $Y_\alpha=0$ or $Y_\alpha=2T_\alpha$)
is given by:
\begin{equation}
G_W^+ = \frac{g}{\sqrt{2} m_W} \sum_\alpha  v_\alpha T^{+,0}_\alpha 
     \phi_{\alpha}^+  \label{Gwphi}   .
\end{equation}

%%%%%%%%%%%%%%%%%%%%%%%%%%%%%%%%%%%%%%%%%%%%%%%%%%%%%%%%%%%%%%%%%%%%%%%%%%
\subsection{\bf The vertex  $H^+_\alpha W^- h^0$}
%%%%%%%%%%%%%%%%%%%%%%%%%%%%%%%%%%%%%%%%%%%%%%%%%%%%%%%%%%%%%%%%%%%%%%%%%

To derive the form of this vertex, we have
to determine the physical charged Higgs states $H_\alpha^+$, 
which must be orthogonal of the Goldstone boson $G_W^+$. For this,
one needs to construct and diagonalize the Higgs mass matrix, 
which requires to study the Higgs potential. However, in this
section we shall procced as general as possible, and we will
only indicate the mixing matrix for the charged and neutral
Higgs states.

Using the previous expansion for $\Phi_\alpha$ (equation (\ref{mul})),
as well
as the properties of the spinors $\chi_\alpha$ 
(equation (\ref{normalization})), and substituting them in the
 equation (\ref{gWHh}),
we  obtain the general expression for the vertices of the type
$W^+ H^{n(+)}  H^{(n+1) (-)}$ as follows
\footnote{A phenomenological study of the double-charged
Higgs vertex is underway \cite{jaimew}.},
\begin{eqnarray}
({\mathcal L}_K)_{W^\mp \phi^\pm \phi^0 } & = &
\frac{ig}{\sqrt{2}} W^-_\mu \sum_\alpha \sum_n T_\alpha^{-,n}
\bigg[ (\phi_\alpha^{(n-1)(+)} )^* \overleftrightarrow{\partial}_\mu  \phi_\alpha^{n(+)}
\bigg]
\nonumber \\
&- & \frac{ig}{\sqrt{2}} W^+_\mu \sum_\alpha \sum_n T_\alpha^{+,n-1}
\bigg[ 
\phi_\alpha^{(n-1)(+)} \overleftrightarrow{\partial}_\mu  (\phi_\alpha^{n(+)})^*  
\bigg] ,
\end{eqnarray}
\ni where $ a\overleftrightarrow{\partial}_\mu b= a\partial_\mu b - b\partial_\mu a $.
 We pick now the terms with $n=0,1$, to obtain the following expression 
for the vertex $W^- \phi^+ \phi^0$, 
\begin{eqnarray}
({\mathcal L}_K)_{W^\mp \phi^\pm \phi^0 } & = &
\frac{ig}{\sqrt{2}} W^-_\mu \sum_\alpha  \bigg[  T_\alpha^{-,0}
(\phi_\alpha^{-} )^* \overleftrightarrow{\partial}_\mu  \phi_\alpha^{0}
 + T_\alpha^{-,1}
 (\phi_\alpha^{0} )^* \overleftrightarrow{\partial}_\mu  \phi_\alpha^{+}
\bigg]
\nonumber \\
&- & \frac{ig}{\sqrt{2}} W^+_\mu \sum_\alpha \bigg[  T_\alpha^{+,-1}
\phi_\alpha^{-}  \overleftrightarrow{\partial}_\mu  (\phi_\alpha^{0})^* 
+  T_\alpha^{+,0}
\phi_\alpha^{0}  \overleftrightarrow{\partial}_\mu  (\phi_\alpha^{+})^*   
\bigg]  .
\end{eqnarray}

If we focus our attention on Higgs multiplets with integer $T_\alpha$ 
and $Y_\alpha= 0$ (or $Y_\alpha = 2T_\alpha$), then  we have  
$ T_\alpha^{-,0}= T_\alpha^{-,1}= T_\alpha^{+,-1}
=  T_\alpha^{+,0}$ 
(or $ T_\alpha^{-,0}= T_\alpha^{+,-1}=0 $ and  $T_\alpha^{-,1}
=  T_\alpha^{+,0} $). Furthermore, for the case   $Y_\alpha= 0$
we use the phase conventions:
 $\phi_{\alpha, T}^+ = - (\phi_{\alpha ,T}^-)^*$.
 Thus,  for theses cases the coupling $W^- \phi^+ \phi^0$, has
the following form:
\begin{equation}
\begin{array}{lcr}
({\mathcal L}_K)_{W^\mp \phi^\pm \varphi^0 } & = &
\displaystyle{\frac{ig}{\sqrt{2}}} W^-_\mu \sum_\alpha  T_\alpha^{+,0}
 \varphi_\alpha^{0} \overleftrightarrow{\partial}_\mu  \phi_\alpha^{+} +
 \mbox{h.c.} \, .
\end{array}
\end{equation}
Since $T_\alpha^{+,0}=const$, the strenght of the vertex will depend on
the mixing factors. Namely,
in order to obtain the  coupling $W^- H^+ h^0$ it is necessary
to determine the physical Higgs states of the charged and 
neutral sector (CP-even). 
Since we have assumed that the Higgs potential is CP-invariant, 
the imaginary and real parts of the neutral scalar fields do not mix.
Thus, the physical  neutral Higgs bosons (CP-even)  are determined 
from $Re \phi^0_\alpha =\varphi^0_\alpha  $.

For the charged Higgs, we define  a unitary rotation that gives the
physical mass eigenstates  $H_\alpha^+$ as: 
\begin{equation}
H_\alpha^+ = \sum_\beta U_{\alpha \beta } \phi^+_\beta \label{rotation} .
\end{equation}
Then, we choose the first charged field $H_1^+$ to be the Goldstone
boson $G_W^+$, while the physical charged fields $H_\alpha^+$ (for $\alpha \geq 2$)  
are orthogonal states to the Goldstone boson. Then, 
we can  fix the first row of the matrix U, through the
following expression
\begin{equation}
U_{1 \beta} = \displaystyle{\frac{g}{\sqrt{2} m_W}} v_\beta T^{+,0}_\beta .
\end{equation}

For the physical neutral Higgs eigenstates $H^0_\beta$, 
one only needs to consider $Re \phi^0_\alpha$, and
we introduce a similar unitary rotation ($V_{\beta \gamma }$) 
that gives the mass-eigenstates, namely:
\begin{equation}
H_\beta^0 = \sum_\beta V_{\beta \gamma} \phi^0_\beta \label{rotation2}  . 
\end{equation}
We can also choose  the first physical neutral field 
$H_1^0$ to be  the lightest Higgs boson, which can be identified as
the light SM-like state preferred by EW radiative
corrections.  

Thus, using these rotations $U$ and $V$, as well as the previous 
conventions, we find that the vertex $W^+ H^-_\alpha H^0_\beta$ 
is given by
\begin{equation}
({\mathcal L}_K)_{W^\mp H^\pm_\alpha H^0_\beta }  = 
\frac{ig}{2} \bigg[ \sum_{\alpha \geq 2,\beta} \eta_{\alpha,\beta}
 H^0_\beta \overleftrightarrow{\partial}_\mu   H^+_\alpha \bigg]  W^-_\mu .   
\label{eta}
\end{equation}
\ni where:
\begin{equation}
\eta_{\alpha,\beta}  =  \sqrt{2}  \sum_{\gamma}
T_\gamma^{+,0} V_{\beta\gamma}^* U_{\alpha \gamma}^* ;
\end{equation}
\ni it depends on the quantum number $T_\alpha$ and the mixing matrices
$U \& V$, but not on the vev's.

%%%%%%%%%%%%%%%%%%%%%%%%%%%%%%%%%%%%%%%%%%%%%%%%%%%%%%%%%%%%%%%%%%%%%%%%
\subsection{Application: a model with one Higgs doublet and
one real triplet}
%%%%%%%%%%%%%%%%%%%%%%%%%%%%%%%%%%%%%%%%%%%%%%%%%%%%%%%%%%%%%%%%%%%%%%%%

To illustrate the above formulae, we can write the
corresponding expressions for a model that includes
a complex Higgs doublet, i.e. $T=1/2$, $T_3=\pm 1/2$, 
and a real triplet, with $ T=1$, $Y=0$. 
In matrix representations, the eigenstates of $T$ 
for the doublet are given by 
\begin{equation}
\begin{array}{rcl}
 | \displaystyle{\frac{1}{2}}, \displaystyle{\frac{1}{2}} > & = & \left( \begin{array}{c}
1\\
0
\end{array}\right)= \chi^+  , \qquad
| \displaystyle{\frac{1}{2}}, -\displaystyle{\frac{1}{2}} >= \left( \begin{array}{c}
0\\
1
\end{array}\right)= \chi^0 , \\
 T_3 & = & \displaystyle{\frac{1}{2}}
\left( \begin{array}{cc}
1  & 0 \\
0 &  -1
\end{array}
\right) .
\end{array}
\end{equation}
Then, we can expand the Higgs doublet $\Phi_D$ in terms
of the spinors $\chi^{+,0}$ as follows: 
\begin{equation}
\Phi_{D} = \left( \begin{array}{c}
\phi_{D}^{+}\\
\phi_{D}^{0}
\end{array}\right) = \phi_{D}^{+} \chi^+ 
+ \phi_{D}^{0} \chi^0 = \sum_{n=0}^1 \phi_{D}^{n(+) } \chi^{n(+) } .
\end{equation}
For this representation the operator $T^\pm$ are
\begin{equation}
T^+=
\left( \begin{array}{cc}
0  & 1 \\
0 &  0
\end{array}\right) , \qquad
T^-= (T^+)^\dag =
\left( \begin{array}{cc}
0  & 0 \\
1 &  0
\end{array}\right) .
\end{equation}
And their action on the spinors $\chi^{+,0}$ are: 
$$T^+ \chi^+ = 0 , \,\,\,
T^+ \chi^0 = \chi^+ , \,\,\, 
 T^- \chi^+ = \chi^0 , \,\,\,
T^- \chi^0 = 0  .$$
\noindent On the other hand, for a Higgs triplet with  $T=1$,  one has
$T_3=1, 0, -1 $ and the spinors associated with the isospin
eigenstates are given by:
\begin{equation}
 | 1, 1 >= \left(\begin{array}{c}
1\\
0 \\
0
\end{array}\right)= \chi^+ ,
\,\,\,
|1, 0>= \left(\begin{array}{c}
0\\
1 \\
0
\end{array}\right)= \chi^0 ,
\,\,\,
|1,-1>=\left(
\begin{array}{c}
0\\
0 \\
1
\end{array}
\right)= \chi^- \,\, . 
\end{equation}
\noindent $T_3$ has the  matrix representation 
\begin{equation}
T_3=
\left( \begin{array}{ccc}
1  & 0 & 0 \\
0  & 0 & 0 \\
0 &  0 & -1
\end{array}
\right) .
\end{equation}

 Thus, the Higgs triplet  $\Phi_T$ with hypercharge $Y=0$, 
can be expanded in terms of the spinors  
$\chi^{+,0,-}$ as follows
\begin{equation}
\Phi_{\alpha} = \left( \begin{array}{c}
\phi_{\alpha}^{+}\\
\phi_{\alpha}^{0} \\
\phi_{\alpha}^{-}
\end{array}\right) = \phi_{\alpha}^{+} \chi^+ 
+ \phi_{\alpha}^{o} \chi^0 + \phi_{\alpha}^{-} \chi^- = \sum_{n=-1}^1 
\phi_{\alpha}^{n(+) } \chi^{n(+) } .
\end{equation}
In this case, the action of matrix  $T^{\pm}$ on the spinors $\chi^{+,0,-}$ 
are given by
$ T^+ \chi^+ = 0 , \,
  T^+ \chi^0 = \sqrt{2} \chi^+ , \, 
  T^+ \chi^- = \sqrt{2}\chi^0 , \,
  T^- \chi^+ = \sqrt{2} \chi^0 , \,
  T^- \chi^0 = \sqrt{2} \chi^- , \, T^- \chi^- = 0$.
The model includes one charged Higgs and two neutral CP-even states,
which are obtained from the weak eigenstates by orthogonal 
$2 \times 2$ rotations, which are parameterized by mixing angles $\alpha$ and
$\delta$, respectively. Therefore, one can write the factor
$\eta$ in terms of these mixing angles; 
for the light state it goes
as follows:
\begin{equation}
\eta_{h^0}=\cos \gamma \sin \delta + \sqrt{2} \sin \gamma \cos \delta .
\end{equation}
\ni Similarly, for the heavier neutral Higgs we obtain:
\begin{equation}
\eta_{H^0}= \sin \gamma \sin \delta + \sqrt{2} \cos \gamma \cos \delta ,
\end{equation}
\ni where $\tan 2\gamma$ depends of parameters of the Higgs potential.

Thus, the coupling $H^+W^-h^0$ is quite sensitive to the structure of the
covariant derivative, and could be one place where to look
for deviations from the minimal THDM (or SUSY) prediction at tree-level.

%%%%%%%%%%%%%%%%%%%%%%%%%%%%%%%%%%%%%%%%%%%%%%%%%%%%%%%%%%%%%%%%%%%%%%%
\section{The vertex $H^+W^-h^0$ in the THDM and MSSM}
%%%%%%%%%%%%%%%%%%%%%%%%%%%%%%%%%%%%%%%%%%%%%%%%%%%%%%%%%%%%%%%%%%%%%%%

One of the simplest models that predicts a charged Higgs is the THDM,
which includes two scalar doublets of equal hypercharge, namely
$\Phi_1=(\phi^+_1,\phi^0_1)$ and $\Phi_2=(\phi^+_2,\phi^0_2)$,
this is indeed the Higgs content used in the minimal SUSY
extension of the SM (MSSM). Besides the charged Higgs ($H^\pm$),
the spectrum of the THDM includes two neutral CP-even
states ($h^0,H^0$, with $m_{h^0} < m_{H^0}$), as well as a neutral CP-odd
state ($A^0$). 

Diagonalization of the charged mass matrices gives the expression for
the charged Higgs mass-eigenstate:
$H^+=\cos\beta \phi^+_1 + \sin\beta \phi^+_2$,
where $\tan\beta(=v_2/v_1)$ denotes the ratio of v.e.v.'s
from each doublet. 
In this case the factor $\eta$ that appears in the vertex 
$H^+W^-h^0$ is given by:
\begin{equation}
\begin{array}{rcl}
\eta_{h^0} & = &  \sin \beta \sin \alpha + \cos \beta \cos \alpha \\
           & = & \cos (\beta-\alpha) .
\end{array}
\end{equation}
\ni Similarly, for the heavier neutral Higgs we obtain:
$\eta_{H^0} = \sin (\beta-\alpha)$. In these cases it is possible for the
vertices $H^+ W^-h^0(H^0)$ to vanish, but only thanks to ad hoc combination
of mixing angles.

%%%%%%%%%%%%%%%%%%%%%%%%%%%%%%%%%%%%%%%%%%%%%%%%%%%%%%%%%%%%%%%%%%%%%%%
\subsection{The decay $H^+W^-h^0$ in the THDM }
%%%%%%%%%%%%%%%%%%%%%%%%%%%%%%%%%%%%%%%%%%%%%%%%%%%%%%%%%%%%%%%%%%%%%%%

The vertex $H^+W^-h^0$ could induce the decay $H^+ \to W^+h^0(H^0)$, whenever 
it is kinematically allowed. Since the charged and neutral Higgs masses are 
given by:
\begin{equation}
\begin{array}{rcl}
m^2_{H^{\pm}}&=& m^2_A + \dis{\frac{2m^2_W}{g^2}} (\lambda_5-\lambda_4) \\
   m^2_{h^0} &=& m^2_A c^2_{\beta-\alpha} + v^2 \left[ \lambda_1 c^2_\beta s^2_\alpha +
                 \lambda_2 s^2_\beta c^2_\alpha -2 \lambda_T c_\alpha c_\beta
                 s_\alpha s_\beta + \lambda_5 c^2_{\beta-\alpha} \right]  ,   \\
   m^2_{H^0} &=& m^2_A s^2_{\beta-\alpha} + v^2 \left[ \lambda_1
                 c^2_\beta c^2_\alpha + \lambda_2 s^2_\beta s^2_\alpha +
                 2 \lambda_T c_\alpha c_\beta s_\alpha s_\beta +
                 \lambda_5 s^2_{\beta-\alpha} \right]  ,
\end{array}
\end{equation}
\ni where $\lambda_i$ are the parameters of the quartic terms that appear in
the Higgs potential, $\lambda_T=\lambda_3+\lambda_4+\lambda_5$ \cite{jfg,npb380}.
Therefore, one can write the following conditions on the Higgs parameters
for the decay $H^+ \rightarrow W^+ h^0$ to procced:
\begin{equation}
{\cos}^2 \beta \lambda_1+ {\sin}^2 \beta \lambda_2 \leq
- \dis{\frac{2}{v^2}}  \left( m^2_W - \dis{\frac{\mu^2_{12}}{\cos \beta
\sin \beta}} \right) - \lambda_4  - \lambda_5
\end{equation}
In the decoupling limit approximation, where
$\alpha \simeq  \beta - \frac{\pi}{2}$
and $\mu_{12}^2=0$ (with $\lambda_6=\lambda_7=0$)
this condition becomes
\begin{equation}
{\cos}^2 \beta \lambda_1+ {\sin}^2 \beta \lambda_2 \leq
- \dis{\frac{2 m^2_W}{v^2}} - \lambda_4  - \lambda_5
\end{equation}
\ni which also corresponds to the case when the scalar potential respects
a discrete symmetry under which one doublet changes sign.
Thus we see that there are regions of parameters where the decay
$H^+ \to W^+h^0$ can proceed. The corresponding decay width is given by:
\begin{equation}
\begin{array}{rcl}
\Gamma(H^+ \to W^+ h^0) & = & 
\displaystyle{\frac{g^{2}\lambda^{\frac{1}{2}} (m^2_{H^+},m^2_{W},
m^2_{h^0}) }
{64 \pi m^{3}_{H^+}}} {\mid \eta_{h^0} \mid}^2 \nonumber \\
&\times& \left[ m_{W}^2- 2(m_{H^{+}}^2 + m^2_{h^0}) +
\displaystyle{\frac{(m_{H^{+}}^2- m^2_{h^0})^2}{m^2_{W}}} \right]
\end{array}
\end{equation}
\ni where $\lambda$ is the usual kinematic factor,
$ \lambda(a,b,c)= (a-b-c)^{2}-4bc$.
This decay  mode has been studied in the literature \cite{hcwhdetect},
and it is concluded that the coming large hadron
collider (LHC) can detect it.
 For the light SM-like Higgs, this decay is proportional to the factor
$\eta^2_{h^0}= \cos^2 (\beta -\alpha )$, which will determine its strength.

Other relevant decays of the charged Higgs boson are the modes into
fermion pairs, which include the decays $H^+ \to \tau\nu_\tau, c \bar{b}$,
and possibly into $t\bar{b}$. If the
charged Higgs is indeed associated with the Higgs mechanism, its
couplings to fermions should come from the Yukawa sector, and the
corresponding decays should have a larger BR for the modes involving the
heavier fermions. A very simple test of this could be
done through a comparison of the modes $H^+ \to \tau\nu_\tau$ and
$H^+ \to \mu\nu_\mu$, which should have very different BR's.

To evaluate the branching ratios within the THDM we have used the expressions for the decay
widths of the tree-level modes, as appearing in ref. \cite{kanehunt}.
We  take $m_t=175\;$ GeV, 
and the values for the electroweak parameters of the
Table of Particle Properties \cite{Partdat}.
 We shall only comment on the resulting BR for
the charged Higgs for the following scenarios, 
and assume here $m_h=115$ GeV. A more detailed discussion
of the THDM case was presented in ref. \cite{ourpaper}.

{\bf a) Non-decoupling scenario-A.} We consider here a large mass difference 
between $A^0$ and $H^+$, i.e. $m_{H^+}- m_A=300$ GeV, with 
$m_H \simeq m_{H^+}$, and also $\alpha\simeq \beta-\pi/2$. 
In this case we find that the mode $W^+h^0$ has a BR about
$10^{-3}$ ($2\times 10^{-5}$), for $\tan\beta=7$ (30)
and $H^+=300$ GeV.    
On the other hand, the BR for the radiative modes $H^+ \to W^+Z$ and
$H^+ \to W^+\gamma$ is about 
$4 \times 10^{-2}$ ($4 \times 10^{-3}$), 
for  is about $2 \times 10^{-6}$ 
($2\times 10^{-7}$), respectively.
Thus, for this case, $H^+\to W^+Z$ dominates. 

{\bf b) Non-decoupling scenario-B.} Here we also assume a large 
mass difference between $A^0$ and $H^+$, i.e. $m_A-m_{H^+}=300$ GeV, 
with $m_H \simeq m_{H^+}$,
but now with $\alpha \simeq \beta-\pi/4$.
In this case we find that the BR for 
the mode $W^+h^0$ has a BR about
$1$ ($0.2$) for $\tan \beta=7$ (30) and $H^+=300$ GeV.
Similarly, the BR for the mode $H^+ \to W^+Z$ is about
$ 10^{-2}$ ($4 \times 10^{-3}$), whereas
the BR for $H^+ \to W^+\gamma$ is about
$4\times 10^{-7}$ ($ 2\times 10^{-7}$).
 In this scenario $H^+ \to W^+h^0$ clearly dominates, 
even above the $t\bar{b}$ mode.

%%%%%%%%%%%%%%%%%%%%%%%%%%%%%%%%%%%%%%%%%%%%%%%%%%%%%%%%%%%%%%%%%%%%%%%
\subsection{The decay $H^+W^-h^0$ in the  MSSM with radiative corrections }
%%%%%%%%%%%%%%%%%%%%%%%%%%%%%%%%%%%%%%%%%%%%%%%%%%%%%%%%%%%%%%%%%%%%%%%

As it was mentioned before, one of the most popular motivations 
for the THDM, is that such Higgs sector 
is in fact the one of the minimal SUSY extension of the SM (MSSM). 
The masses of the two CP-even neutral Higgses ($h^0,H^0$)
and the charged pair ($H^\pm$), are conveniently determined in terms of
the mass of the CP-odd state ($A^0$) and $\tan\beta=v_2/v_1$.
The Higgs potential of the MSSM has less free parameters than the
THDM; in particular the quartic couplings are given in terms of
the gauge couplings, which then implies that the light Higgs boson
must satisfy the (tree-level) bound $m_{h^0} \leq \cos 2 \beta \, m_Z$.
However this relation receives important corrections form top/stop 
loops, which give an aproximate bound $m_{h^0} \leq 130$ GeV \cite{cotaneu}.

In the decoupling limit ($m_A >> m_Z$) the parameters of the potential give the approximate
relation:  $c_{\beta -\alpha }^2 \sim m_z^2/m_{A^0}^2 $,
which tends to be small for large values of $m_{A^0}$.
One also obtains an approximately degenerated spectrum of heavy Higgs bosons, 
i.e. $m_{H^+} \simeq m_{H^0} \simeq m_{A^0}$, while the mixing angles
satisfy the approximate relation: $\alpha \simeq \beta-\pi/2$.
Therefore, only the decay mode $W^+ h^0$ is allowed for most regions of 
parameter space of the MSSM. One obtains a typical BR of the order 
$2\times 10^{-2}$ ($ 7 \times 10^{-5}$) for $m_{H^+}=300$ GeV and 
$\tan \beta = 7 \, (30)$.

We have performed a detailed parametric search for contour
regions for the branching ratio of $H^\pm \to W^\pm +h^0$, using the
program HDECAY \cite{hdecay}, and the results are shown in fig. 1.
We thus see that the BR is larger for
small values of $\tan\beta$ and almost independent of $m_{A^0}$.

%%%%%%%%%%%%%%%%%%%%%%%%%%%%%%%%%%%%%%%%%%%%%%%%%%%%%%%%%%%%%%%%%%%%%%%
\section{The vertex $H^+W^-h^0$ in a SUSY models with a Higgs triplet}
%%%%%%%%%%%%%%%%%%%%%%%%%%%%%%%%%%%%%%%%%%%%%%%%%%%%%%%%%%%%%%%%%%%%%%%

The supersymmetric model with two doublets and a complex triplet is one of
the simplest extension of the minimal supersymmetric model that allows to
study phenomenological consequences of an explicit breaking of the
custodial symmetry SU(2) \cite{ESQuia}. 

%%%%%%%%%%%%%%%%%%%%%%%%%%%%%%%%%%%%%%%%%%%%%%%%%%%%%%%%%%%%%%%%%%%%%%%%%%%
\subsection{The Higgs sector of the model }
%%%%%%%%%%%%%%%%%%%%%%%%%%%%%%%%%%%%%%%%%%%%%%%%%%%%%%%%%%%%%%%%%%%%%%%%%%

The model includes two Higgs 
doublets and a (complex) Higgs triplet given by
\begin{equation}
\Phi_1 = \left( 
\begin{array}{c}
{\phi_1}^0 \\ 
{\phi_1}^-
\end{array}
\right) \,\,\, , \,\,\, \Phi_2= \left( 
\begin{array}{c}
{\phi_2}^+ \\ 
{\phi_2}^0
\end{array}
\right) \,\,\, , \,\,\,
\sum = \left( 
\begin{array}{cc}
\sqrt{\frac{1}{2}} \xi^0 & - \xi_2^+ \\ 
\xi_1^- & - \sqrt{\frac{1}{2}} \xi^0
\end{array}
\right) \,\,\, .
\end{equation}
The Higgs triplet is described in terms of the $2 \times 2$ matrix representation; $\xi^0$ is the complex neutral field, and $(\xi_1^-)^\ast, (\xi_2^+)$
denote the charged scalars. 
The most general gauge invariant and renormalizable superpotential, that can
be written for the Higgs superfields $\Phi _{1,2}$ and $\Sigma $ is given
by: 
\begin{equation}
W=\lambda \Phi _{1}\cdot \Sigma \Phi _{2}+\mu _{D}\Phi _{1}\cdot \Phi
_{2}+\mu _{T}\mbox{Tr}(\Sigma ^{2})\,\,\,,
\end{equation}
where we have used the notation $\Phi _{1}\cdot \Phi _{2}\equiv \epsilon
_{ab}\Phi _{1}^{a}\Phi _{2}^{b}$.
The resulting scalar potential involving only the Higgs fields 
is then written as 
$$
V=V_{SB}+V_{F}+V_{D}\,\,\,, 
$$
\noindent where $V_{SB}$ denotes the most general soft-supersymmetry breaking
potential \cite{TRIP}. In turn, the full scalar potential 
can be splitted into its neutral and charged parts, 
i.e. $V = V_{charged} + V_{neutral} $.

From the Higgs potential one derives the minimization conditions and the 
scalar mass matrices. For the neutral  scalars we have 
that the resulting mass matrix splits into two blocks, one of them 
(the imaginary components) is associated with the pseudoscalar 
Higgs states, while the other one (real components) describes the 
masses of the scalar Higgs bosons. The mass matrix for the imaginary parts
contains a massless state, which is the Goldstone boson $G^{0}$ that 
gives mass to the Z boson. Whereas the mass matrix for charged states 
include also a massless state $G^{+}$, which give  mass to $W^+$ boson 
($G^{+ \, *} \equiv G^-$).

Besides the supersymmetry-breaking mass
terms, $m_i^2$ ($i = 1,\,2,\,3$), the potential depends on the parameters
$\lambda, \,\, \mu_D, \,\, \mu_T, \,\, A, \,\, B$. For simplicity we shall
assume that there is no CP violation in the Higgs sector, and thus all
the parameters and the v.e.v.'s are assumed to be real. The explicit 
expressions of the Higgs potential are given in ref \cite{TRIP}.

We can also combine the v.e.v.'s of the Higgs doublet as 
$v_{D}^{2}\equiv v_{1}^{2}+v_{2}^{2}$
and define $\mbox{tan}\beta \equiv {v_{2}}/{v_{1}}$. 
Further, the relations between ($v_{D},\,\,v_{T}$) 
and ($m_{W}^{2},\,\,m_{Z}^{2}$) are 
\begin{equation}
\begin{array}{cl}
m_{W}^{2}= & \frac{1}{2}g^{2}(v_{D}^{2}+4v_{T}^{2}), \\ 
m_{Z}^{2}= & {\displaystyle{\frac{{\frac{1}{2}g^{2}v_{D}^{2}}}{{\mbox{cos}
^{2}{\theta }_{W}}}}}\,\,\,.
\end{array}
\end{equation}
which imply that the tree-level $\rho $-parameter is different from one,
namely, 
\begin{equation}
\rho \equiv \frac{M_{W}^{2}}{M_{Z}^{2}\mbox{cos}^{2}\theta _{W}}
=1+4R^{2},\,\,\,\,\,\,R\equiv \frac{v_{T}}{v_{D}}\,\,\,.
\end{equation}
The bound on $R$ is obtained from the $\rho $ parameter, which lays in the range 0.9799-1.0066 at $95\,\%$ c.l., thus,  
$R \leq 0.04 \,\, (95\,\%\,\,\mbox{c.l.})$
and then $v_{T}\leq 9\,\,\mbox{GeV}$, at 95 \% c.l. \cite{TRizzo}.
This bound must be respected in our numerical analysis.

%%%%%%%%%%%%%%%%%%%%%%%%%%%%%%%%%%%%%%%%%%%%%%%%%%%%%%%%%%%%%%%%%%%%%%%%
\subsection{Mass spectrum}
%%%%%%%%%%%%%%%%%%%%%%%%%%%%%%%%%%%%%%%%%%%%%%%%%%%%%%%%%%%%%%%%%%%%%%%%%

The diagonalization of the mass matrices, and the resulting mass eigenvalues
and mixing matrix, will allows us to analyze the coupling
$ H^+_\alpha W^{-}h^0_\beta$ ($\alpha=1,2,3$ and $\beta=1,2,3$).
The CP-even mass eigenstates are denoted by $h^0$, $H^0_{1}$ y $H^{0}_{2}$,
ordered according to their masses, $ m_{h^0} < m_{H^{0}_{1}} < m_{H^{0}_{2}}$.
The charged Higgs states are denoted by $H^{\pm}_{\alpha}$
with $ m_{H^{\pm}_{1}}< m_{H^{\pm}_{2}} < m_{H^{\pm}_{3}}$. Because of the 
large number of parameters appearing in the model, which include 
$\tan \beta ,R,\lambda ,\mu _{D},\mu _{T},A,B_{D},B_{T}$, one has to consider
 some simplified cases, for which we shall try to identify usefull relations 
or trends for the behaviour of Higgs masses and couplings. For 
our numerical analysis of the allowed region and Higgs masses, we shall 
consider: a) $\tan \beta $ as an independent-variable; b) $R$ will take the 
representative value 0.01; c) the parameter $\lambda$ will take the values 
0.5, 1.0; and d) the remmaining parameters will cover the ranges allowed by SUSY,
namely masses in the range between 0 and 1000 GeV. Furthermore, we shall 
analyze the following specific scenarios (which were defined and studied in
ref. \cite{TRIP}):

{\bf Scenario I\/:} 
$B_{D}=\mu _{D}=0$, which represent the
scenario when SSB is dominated by the effects of the Higgs triplets, here we
shall also consider the following cases
\begin{description}
\item [A)]$\,B_{T}=\,\, \mu _{T}=\,\,A$
\item [B)]$\,B_{T}=\,\, \mu _{T}=-A$
\item [C)]$\,B_{T}=-\mu _{T}=\,\,A$
\item [D)]$-B_{T}=\,\, \mu _{T}=\,\,A $
\end{description}
{\bf Scenario II\/:}
$\ B_{T}=\mu _{T}=0$, which represent the
scenario when SSB is dominated by the effects of the Higgs doublets; now the
following cases will be considered:
\begin{description}
\item [A)]$\,B_{D}=\,\,\mu _{D}= \,\,A$
\item [B)]$\,B_{D}=\,\,\mu _{D}=-A$
\item [C)]$\,B_{D}=-\mu _{D}=\,\,A$
\item [D)]$-B_{D}=\mu _{D}=\,\,A$
\end{description}
{\bf Scenario III\/:} 
$\left| \ B_{D}\right| =\left| B_{T}\right| =\left| \mu _{D}\right| =\left| \mu _{T}\right| =\left| A\right| $,
both doublets and the triplet contribute to SSB. Within this scenario we shall
consider several cases; for instance A) $B_{D}= B_{T}=\mu _{D}=\mu
_{T}=A$, as well as 15 others combinations with positive
and negative signs.

Then, for each point in parameter space, within the above scenarios, we shall 
determine the allowed regions, by requiring the scalar squared mass 
eigenvalues to be positive, and the Higgs potential laying in a global 
minimum. In these allowed regions  the mases of the physical Higgs 
states of the model are computed numerically.  

%%%%%%%%%%%%%%%%%%%%%%%%%%%%%%%%%%%%%%%%%%%%%%%%%%%%%%%%%%%%%%%%%%%%%%%%%
\subsection{The vertex $W^{+}H^{-}_{\alpha}h^0$ and $W^{+}H^{-}_{\alpha}Z$}
%%%%%%%%%%%%%%%%%%%%%%%%%%%%%%%%%%%%%%%%%%%%%%%%%%%%%%%%%%%%%%%%%%%%%%%%%

We shall apply now the general expression for the vertex
$W^{+}H^{-}_{\alpha}h^0$
derived in sect. 2, for the present SUSY model with a Higgs triplet;
we shall only consider the case of the lightest neutral CP-even scalar.
To present a complete study of the branching ratios of the charged Higgs, 
we shall also discuss the vertex $W^{+}H^{-}_{\alpha}Z$, which could
dominate in some specific scenarios.

Substituting the expression for the rotation matrices of the
charged and neutral Higgs sectors, $U$ and $V$, in the expression
for $\eta_{\alpha,\beta}$ (equation (\ref{eta}))
allows us to
derive the following expression for the coefficient of
the vertex $W^{-} H^+_\alpha h^0$,
namely,
\begin{equation}
\eta_{h^{0}}= \left( \dis{\frac{1}{\sqrt{2}}}V_{11}( U_{2i} - U_{1i})+ 
\dis{\frac{1}{4}}V_{31}( U_{4i} - U_{3i}) \right) \,\,\, ,
\end{equation}
\noindent where $H^{+}_{\alpha}$ denote the charged Higgs bosons of the model,
and $h^0$ corresponds to the lightest scalar Higgs boson of the model.
The terms $U_{ji}$ denote the coefficients in the expansion for
$H^{+}_\alpha$, which is given by:
\begin{equation}
H^{\pm}_\alpha = U_{1i} H_2^+ + U_{2i} H_1^- + U_{3i} \xi_2^+ + U_{4i} \xi_1^{-\,*}
\label{expansion}
\end{equation}
\ni 
while $V_{ij}$ denote the elements of the rotation matrix for
the CP-even neutral sector.

On the other hand, in this model the vertex $H^+_\alpha W^- Z$ is also induced at 
tree level because of the violation of the custodial symmetry.
The expression for the vertex $H^+_\alpha W^- Z$ is given by
\begin{equation}
H_{\alpha}^{+}W^{-}Z: i \,e^2 v_T (U_{3i} - U_{4i})\displaystyle{\frac{\mbox{cos}{\theta}_W}{{\mbox{sin}}^2{\theta}_W}} \,\,\, .
\end{equation}
\noindent  One can see than only the triplet components contribute to 
this vertex, while the dependence on $v_T$ gives a suppression effect. 
In what follows, the coefficients $U,V$, 
are calculated at tree level. 

%%%%%%%%%%%%%%%%%%%%%%%%%%%%%%%%%%%%%%%%%%%%%%%%%%%%%%%%%%%%%%%%%%%%%%%%%%%%
\subsection{Branching ratios for the modes $H^+_\alpha \rightarrow W^+ Z$ and 
$H^+_\alpha \rightarrow W^{+}h^0$}
%%%%%%%%%%%%%%%%%%%%%%%%%%%%%%%%%%%%%%%%%%%%%%%%%%%%%%%%%%%%%%%%%%%%%%%%%

We now discuss the BR for the charged Higgs,
including the decay
widths of the dominant modes of $H^+_\alpha$, which turn out
to be the following modes: 
1)$H^{+}_{\alpha} \rightarrow W^{+}Z$; 
2)$H^{+}_\alpha \rightarrow W^{+}h^0$;
3) $H^{+}_\alpha \rightarrow \tau \nu_{\tau}$; and 
4)$H^{+}_\alpha \rightarrow t \bar{b}$. 
The decay width for each of the above modes  is:
\begin{enumerate}
\item The decay $H^+_\alpha  \rightarrow W^+ h^0$:
\begin{equation}
\begin{array}{cl}
\Gamma \left( H^{+}_\alpha \rightarrow W^{+}h^0 \right) = &
\dis{\frac{g^2 {\lambda}^{1/2}(m_{H_{\alpha}^{+}}^{2},m_{W}^{2},m_{h^0}^{2})}
{64 \pi m_{H_{\alpha}^{+}}^{3}}} {\mid \eta_{h^{0}} \mid}^2  \\
 & \times \[ m_{W}^{2}- 2 (m_{H_{\alpha}^{+}}^{2} + m_{h^0}^{2}) + 
 \dis{\frac{{(m_{H_{\alpha}^{+}}^{2} + m_{h^0}^{2})}^{2}}{m_{W}^{2}}} \]
\end{array}
\end{equation}
\ni where $\lambda^{1/2}$ is the usual kinematic factor  
$\lambda^{1/2}(a,b,c)= (a-b-c)^2 - 4bc$; this decay is proportional to 
the factor $\eta_{h^{0}}^2$.

\item The decay $H_\alpha^{+} \rightarrow W^{+}Z$:
\begin{equation}
\Gamma \left( H_{\alpha}^{+} \rightarrow W^{+}Z \right) = 
\dis{\frac{m_{H_{\alpha}^+}}{16 \pi}} {\lambda}^{1/2}(1,w,z)
\left[ {\mid M_{LL} \mid}^2 +  {\mid M_{TT} \mid}^2  \right]  \,\,\, .
\end{equation}
\ni Here $w=(\frac{m_W}{m_{H_\alpha^+}})^2$ and  $z=(\frac{m_Z}
{m_{H_\alpha^+}})^2$;
${\mid M_{LL} \mid}^2=\dis{\frac{g^2}{4z}} {\mid (1-w-z) F_Z \mid}^2$ and
${\mid M_{TT} \mid}^2= 2 g^2 w {\mid
F_Z \mid}^2$ are the final
polarization contributions of the gauge bosons.

\item The decay $H_{\alpha}^{+} \rightarrow t \bar{b}$:
\begin{equation}
\begin{array}{c}
\Gamma (H_\alpha^{+} \rightarrow t \bar{b}) = 
\dis{\frac{3g^2}{32 \pi m_{W}^{2} m_{H_\alpha^+}^{3}}} 
\lambda^{1/2} ( m_{H_\alpha^{+}}^3, m_{t}^{2},  m_{b}^{2}) \\  
\times \left[ (m_{H_\alpha^+}^3 - m_{t}^{2}- m_{b}^{2})
(m_{b}^{2} {\tan}^2 \beta \, T_{2} +m_{t}^{2} {\cot}^2 \beta \,
T_{1} ) - 4m_{b}^{2}m_{t}^{2}\, T_1 \, T_2 \right] \,\,\, ,
\end{array} 
\end{equation}
\ni where $T_1$ y $T_2$ depend on the mixing angles that diagonalize
the charged Higgs mass matrix, namely:
$T_1= \cot \beta (\dis{\frac{U_{22}}{\cos \beta}})$ and
$T_2= \tan \beta (\dis{\frac{U_{12}}{\sin \beta}})$.

\item The decay $H_{\alpha}^{+} \rightarrow \tau \nu_{\tau}$:
\begin{equation}
\begin{array}{cl}
\Gamma (H_{\alpha}^{+} \rightarrow \tau \nu_{\tau}) = &
\dis{\frac{g^2}{32 \pi m_{W}^{2} m_{H_{\alpha}^{+}}^3}} \lambda^{1/2}\left( 
m^2_{H_{\alpha}^{+}}, 0,  m_{\tau}^{2} \right) \\  
& \times \, m_{\tau}^{2},\ {\tan}^2 \beta \, T_{2} (m_{H_\alpha^+}^3
- m_{\tau}^{2})
\end{array}
\end{equation}

\end{enumerate}

We have then evaluated numerically the BR for these four 
modes, using the previous expressions. For the numerical analyses, we 
considered the scenarios listed above, which have fixed values of
$\lambda$ and tan $\beta$. 

In scenario I, the calculation are performed for
$\lambda = 0.5,1$ and tan $\beta= 5,10,15$. We considered the cases A and D 
within this scenario, for each of the charged Higgs bosons $H^{+}_{\alpha}$. 
In Figs. 2  we present the BR for case  A,  with $\lambda = 0.5$. In this 
scenario, the decay $H_{\alpha}^{+} \rightarrow W^{+}h^0$ is not allowed
for the lightest charged Higgs boson, thus 
and we only show the results for  $H^{+}_{2}$ and  $H^{+}_{3}$. 
For $H^{+}_{2}$, the second lighter charged Higgs boson, 
we can see that $WZ$ is the dominant mode, which would be a
clear signature of the Higgs triplet. The mode $W^+h^0$ 
reaches an important BR, although it is
smaller than the BR for $t \bar{b}$. For the state $H^{+}_{3}$, 
$W^+Z$ has a BR of the order $10^{-1}$; here the mode
$W^+ h^0$ is dominant, while the mode $tb$ becomes dominant when 
$\tan \beta$ increasses.
In turn, the  mode $\tau \nu_{\tau}$ is the most supressed one, 
and it reaches a maximum BR of order $10^{-3}$ when $\lambda = 0.5$ and
large $\tan \beta$ . 

For $\lambda = 1$ in scenario I (see Fig. 3), the mode $W^+h^0$ is dominant
for $H^{+}_{2}$, whereas  for $H^{+}_{3}$ it becomes dominant when 
$\tan \beta$ increasses. The $W^+Z$ mode gets a BR of the order $10^{-2}$,
this is dominating for $H^{+}_{3}$ boson when $\tan \beta$ is small,
while the BR for the mode $t\overline{b}$ increasses  when $\tan \beta$ increase.
For case D, we notice a different behavior, as it is show in Figs. 4 
and Fig. 5, which show the BR for $\lambda = 0.5$ y $\lambda = 1$. 
Now the mode $W^+ h^0$ mode gets a BR of order $10^{-1}$ for
$H^{+}_{2}$; the same occurs for $H^{+}_{3}$ boson.
When $\lambda$ increase, $W^+h^0$ mode is dominant for 
$H^{+}_{3}$, while $W^+Z$ is dominant for $H^{+}_{2}$. 

In scenario II, which mimics the MSSM, 
Fig. 5 and 6 show the BR's for 
$\lambda = 0.5$ and $\lambda = 1$.
 Now, the mode $t\overline{b}$ becomes  dominant for $H^{+}_{1}$, 
the lightest charged Higgs boson, for any combination of 
of parameters. We also notice that 
the mode $W^+h^0$ has a BR of order of $10^{-2}$. 
On the other hand, the mode $W^+Z$ is dominant for $H^{+}_{2}$ and $H^{+}_{3}$,
for any combination of the parameters.
The mode  $W^+h^0$ mode gets suppressed 
 when  $\tan \beta$ increase.

Finally, for scenario III we considere the case F. Here
both doublets and tripet contribute equally to SSB. 
The results for the BR's are show in Fig. 8 and Fig. 9
for $\lambda = 0.5$ and $\lambda = 1$, respectively.
Again, we find that for $H^{+}_{1}$ the dominant 
mode is $t\overline{b}$. The behavior of $W^+Z$ and $^+h^0$ modes is similar to 
the ones from scenario II. 
We also find that the mode $W^+h^0$ reaches larger BR's 
when $\tan \beta$ is large.

%%%%%%%%%%%%%%%%%%%%%%%%%%%%%%%%%%%%%%%%%%%%%%%%%%%%%%%%%%%%%%%%%%%%%%%
\section{Conclusions}
%%%%%%%%%%%%%%%%%%%%%%%%%%%%%%%%%%%%%%%%%%%%%%%%%%%%%%%%%%%%%%%%%%%%%%%

 We have studied the charged Higgs vertex $H^{+}_\alpha W^{+} h^{0}_\beta$, 
within the context of several extensions of the SM that predict this 
vertex. For a Higgs sector that includes arbitrary Higgs representations, 
we were able to derive the general form of this vertex, i.e. its dependence on the 
isospin and hypercharge of the Higgs multiplet.
Then, we evaluate the strength of this vertex for several specific
models, which include: $i)$ the THDM, both generic and the MSSM
version, and $ii)$ models with additional Higgs triplets, 
for both SUSY and non-SUSY cases. When the decay 
$H^{+}_\alpha \rightarrow W^+ h^0$ is allowed, it can reach a BR that
could be detected at LHC, and would permit
to test the charged Higgs quantum numbers. We can summarize our results in terms
of the following classification for the BR, namely:

\begin{itemize}
\item Dominant: Large BR (when $H^{+}_\alpha \rightarrow W^+ h^0$ is
the dominant mode)
\item Moderate: BR $\sim 10^{-1} - 10^{-2}$
\item Small:     BR $\sim 10^{-2} - 10^{-4}$
\item Negligible: BR $ < 10^{-4}$
\end{itemize}

\begin{table}
$$
\begin{tabular}{|c|c|c|c|}
\hline\hline
BR$\left( H^{+}\rightarrow W^{+}h^{0}\right) $ & THDM$^{(1)}$ & MSSM$^{(2)}$
& $
\begin{tabular}{l}
Triplets$^{(3)}$ \\ 
SUSY-Higgs
\end{tabular}
$ \\ \hline
\multicolumn{1}{|l|}{a) Dominant} & 
\begin{tabular}{l}
$\alpha =\beta -\frac{\pi }{4}$ \\ 
tan $\beta =7$%
\end{tabular}
& not possible & \multicolumn{1}{|l|}{$
\begin{tabular}{l}
Scenario I: \\ 
$m_{H^{+}}=100-550$ GeV \\ 
tan $\beta =5$%
\end{tabular}
$} \\ \hline
\multicolumn{1}{|l|}{b) Moderate} & 
\begin{tabular}{l}
$\alpha =\beta -\frac{\pi }{4}$ \\ 
tan $\beta =30$%
\end{tabular}
& 
\begin{tabular}{l}
$m_{H^{+}}=300$ GeV \\ 
tan $\beta \cong 7$%
\end{tabular}
& \multicolumn{1}{|l|}{
\begin{tabular}{l}
\begin{tabular}{l}
Scenario I: \\ 
$m_{H^{+}}=200-300$ GeV \\ 
tan $\beta =5,10$%
\end{tabular}
\\ 
\begin{tabular}{l}
Scenario II: \\ 
$m_{H^{+}}=200$ GeV \\ 
tan $\beta =5,15$%
\end{tabular}
\\ 
\begin{tabular}{l}
Scenario III: \\ 
$m_{H^{+}}=150-300$ GeV \\ 
tan $\beta =30$%
\end{tabular}
\end{tabular}
} \\ \hline
\multicolumn{1}{|l|}{c) Small} & 
\begin{tabular}{l}
$\alpha =\beta -\frac{\pi }{2}$ \\ 
tan $\beta =7$%
\end{tabular}
& 
\begin{tabular}{l}
$m_{H^{+}}=300$ GeV \\ 
$10<$ tan $\beta \lesssim 25$%
\end{tabular}
& \multicolumn{1}{|l|}{
\begin{tabular}{l}
\begin{tabular}{l}
Scenario I: \\ 
$m_{H^{+}}\simeq 250-400$ GeV \\ 
tan $\beta =10$%
\end{tabular}
\\ 
\begin{tabular}{l}
Scenario III: \\ 
$m_{H^{+}}\simeq 200$ GeV \\ 
tan $\beta =5,15,30$%
\end{tabular}
\end{tabular}
} \\ \hline
\multicolumn{1}{|l|}{d) Negligible} & 
\begin{tabular}{l}
$\alpha =\beta -\frac{\pi }{2}$ \\ 
tan $\beta =30$%
\end{tabular}
& 
\begin{tabular}{l}
$m_{H^{+}}=300$ GeV \\ 
tan $\beta \geq 25$%
\end{tabular}
& \multicolumn{1}{|l|}{
\begin{tabular}{l}
\begin{tabular}{l}
Scenario I: \\ 
$m_{H^{+}}\simeq 280$ GeV \\ 
tan $\beta =15$%
\end{tabular}
\\ 
\begin{tabular}{l}
Scenario II: \\ 
$m_{H^{+}}$ $\simeq 200$ GeV \\ 
tan $\beta =5,15,30$%
\end{tabular}
\\ 
\begin{tabular}{l}
Scenario III: \\ 
$m_{H^{+}}\simeq 200$ GeV \\ 
tan $\beta =5,15,30$%
\end{tabular}
\end{tabular}
} \\ \hline\hline
\end{tabular}
$$
\caption{Classification of BR $(H^{+}\rightarrow W^{+}h^{0})$ according
to the scheeme discussed in the text. (1) For the THDM we take
$m_{H^+}-m_A=300$ GeV, $m_{h^{0}}=115$ GeV. (2) We have
$m_{H^+} \approx m_{A^{0}}$ for most regions of parameters.
(3) We consider the cases with $\lambda = 1$, and the scenarios I, II, III
defined in the text.}
\end{table}

We can appreciate that for each model there area regions or values of
parameters that correspond to one of those scenarios. Therefore,
the observation of the decay $H^{+}\rightarrow W^{+}h^{0}$, as the dominant
mode, would correspond to the THDM or scenario I of the
SUSY triplet case, while the moderate case could arise
either of THDM or MSSM (observation of more Higgs bosons with the
predicted properties would then be needed to descriminate among them),
while the detection of several charged and neutral Higgs bosons would
correspond to a model with more elaborated Higgs sector (such as
Higgs triplets). Then, some results for typical BR's within each model are
shown in Table 1.

%%%%%%%%%%%%%%%%%%%%%%%%%%%%%%%%%%%%%%%%%%%%%%%%%%%%%%%%%%%%%%%%%%%%%%%%%%
Acknowledgments: The work of J.H.S. is suported by Programa de
Consolidaci\'on Institucional-Conacyt (M\'exico). J. L. D. C.
is supported by CONACYT-SNI. We would like to thank the Huejotzingo
Seminar on Theoretical Physics, for inspiration and discussions.
%%%%%%%%%%%%%%%%%%%%%%%%%%%%%%%%%%%%%%%%%%%%%%%%%%%%%%%%%%%%%%%%%%%%%%%%%%%


\newpage

{\large List of Figures}

%\listoffigures
\begin{enumerate}
\item BR $(H^+ \rightarrow W^+ h^0)$ in the MSSM, with
radiative corrections to the Higgs mass as included in HDECAY, with
$m_{\widetilde{q}}=500$ GeV, $\mu = 100$ and $A_0=1500$.

\item Branching Ratios of the charged Higgs bosons $H_{\alpha}^{+}$
in the principal modes for the scenario I, case A, considering
$\lambda = 0.5$.
The kind of lines correspond to the different modes as:
(a) dashed: $H^+_\alpha \rightarrow t \overline{b}$;
(b) dotted: $H_{\alpha}^{+} \rightarrow W^+ Z$;
(c) solid:
$H_{\alpha}^{+} \rightarrow \tau \overline{\nu}_{\tau}$;
and dot-dashed: $H_{\alpha}^{+} \rightarrow W^+ h^0$.
The figure show this modes for each charged Higgs boson,
the first row correspond to the $H_{2}^{+}$ and the
second row to $H_{3}^+$. In the columns is shown the different
results to tan $\beta =5,10,15$.

\item The same of Fig. 2, with $\lambda = 1.0$.

\item Scenario I, case D, with $\lambda=0.5$.

\item The same of Fig. 4, with $\lambda=1.0$.

\item Scenario II, case D, with $\lambda=0.5$. In this scenario, we
considered  $\tan \beta = 5,15,30$.
\item The same of Fig. 6, with $\lambda = 1.0$.
\item Scenario III, case F, with $\lambda = 0.5.$.
\item The same of Fig. 8, with $\lambda = 1.0$.
\end{enumerate}

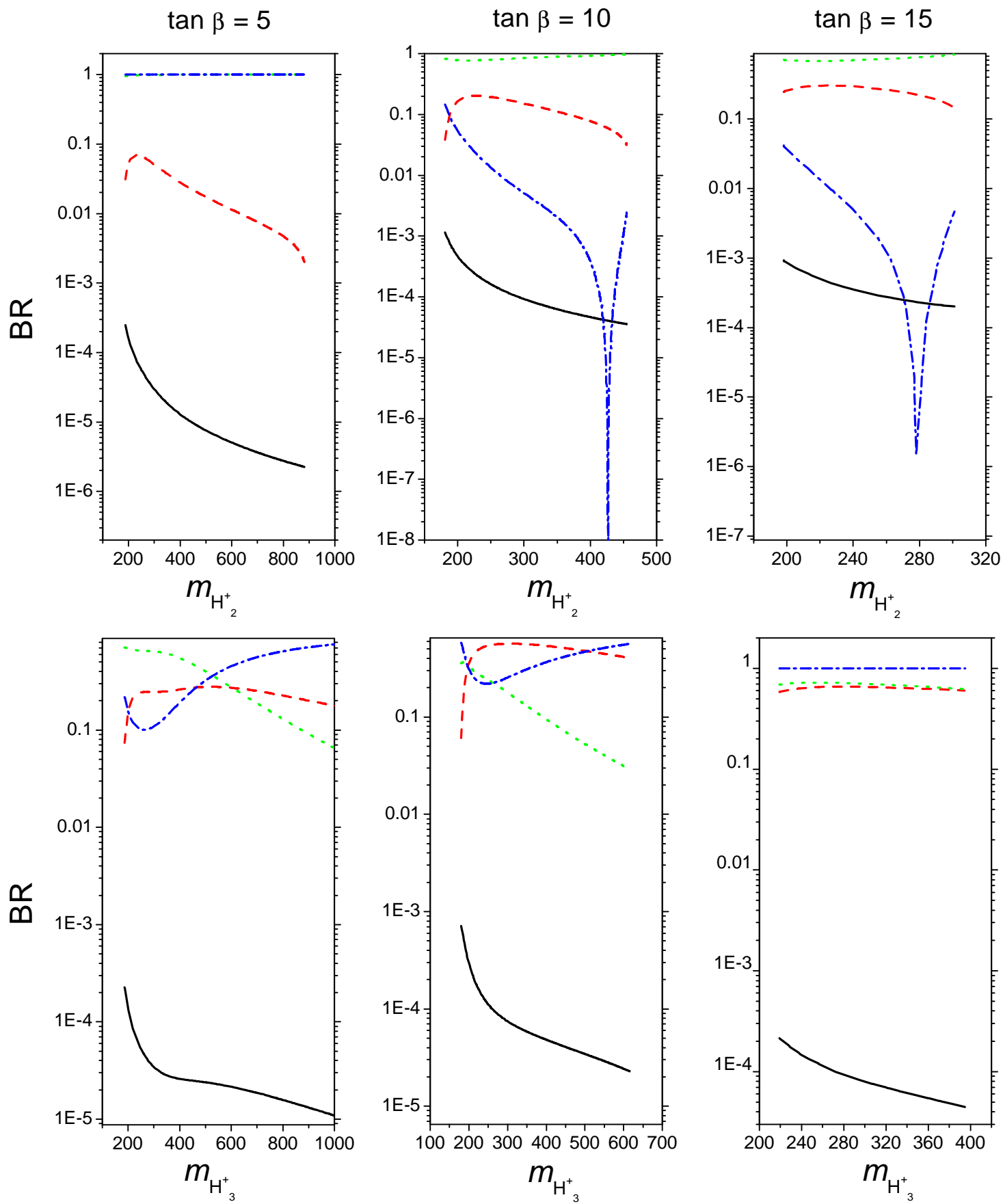
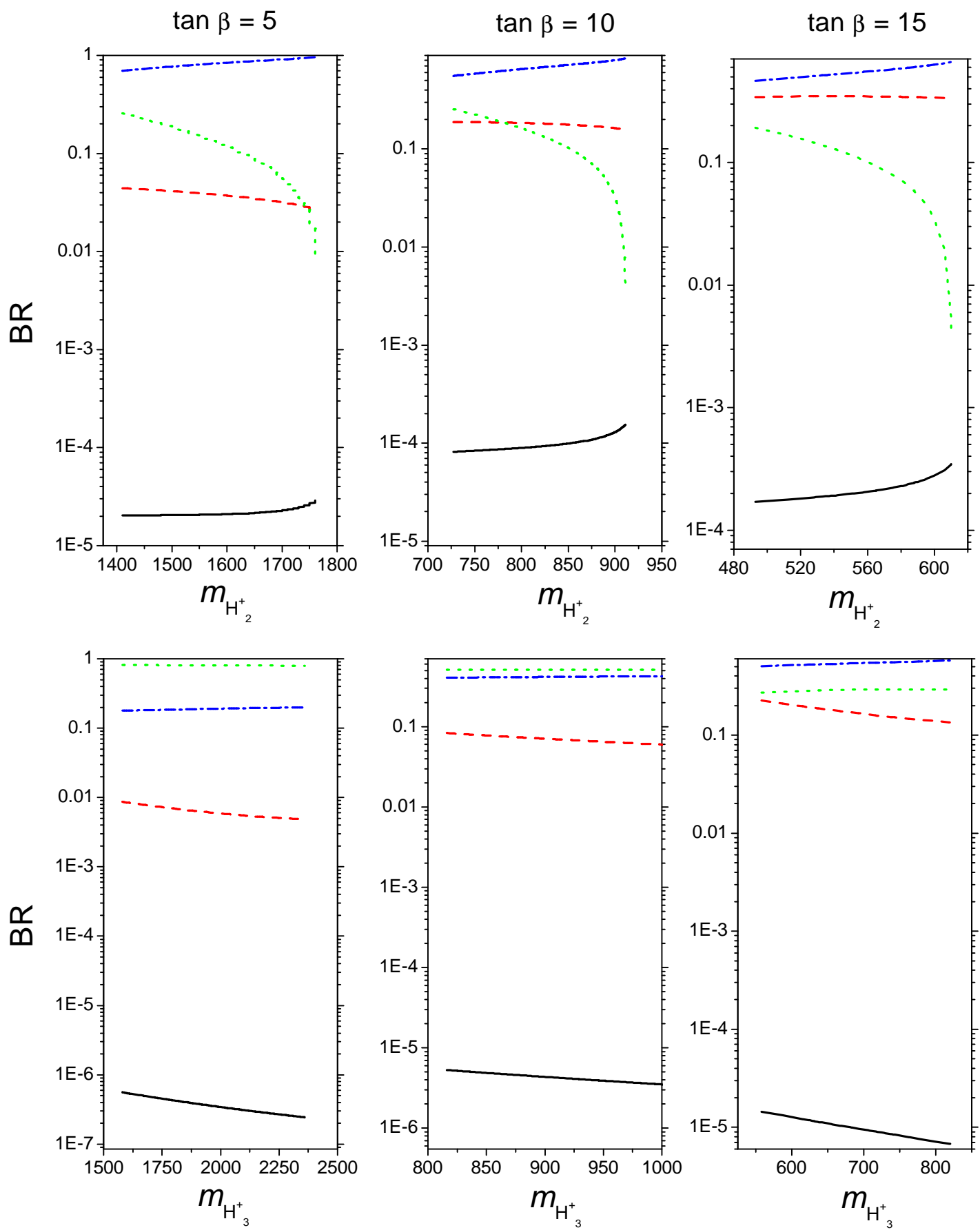
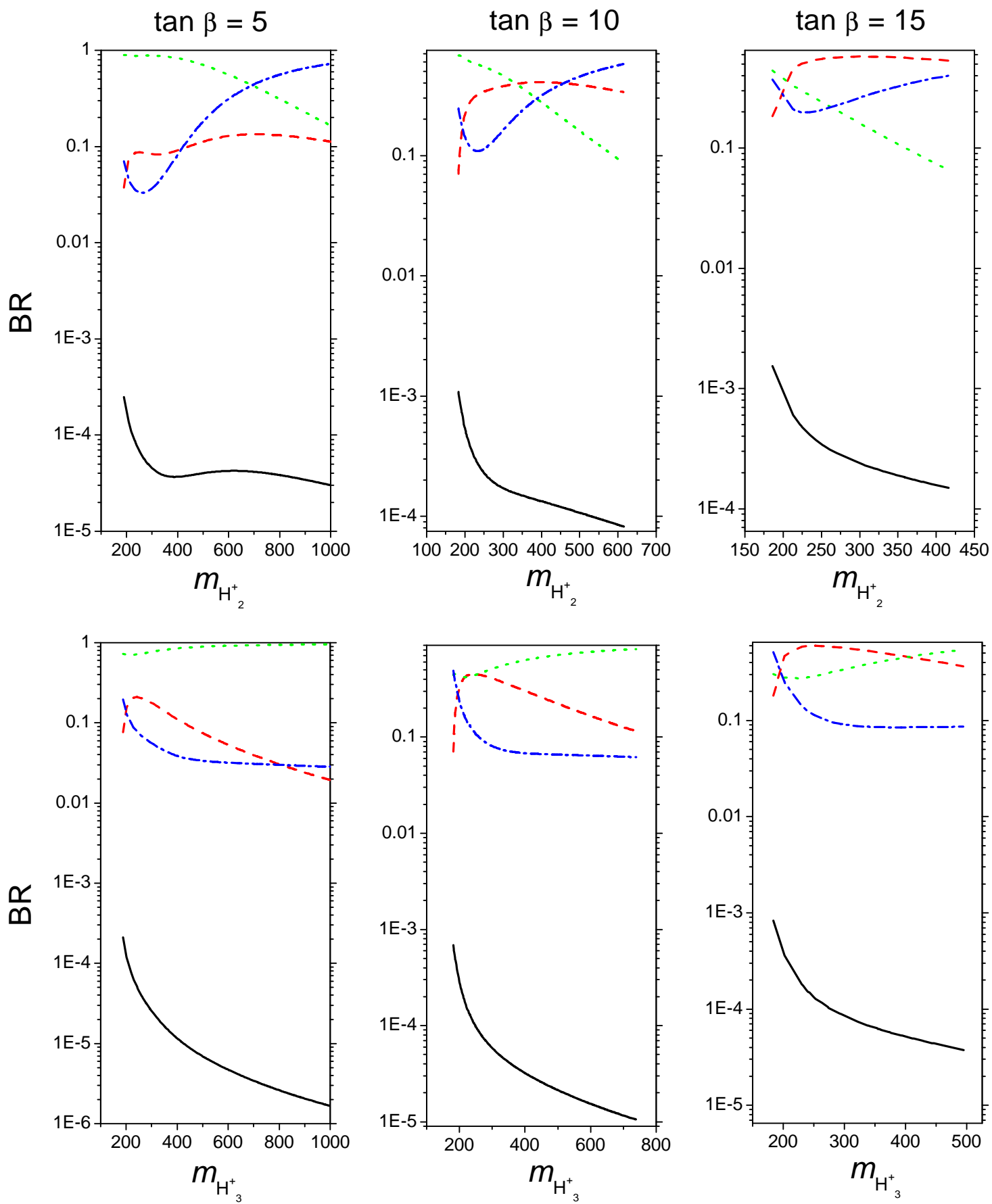
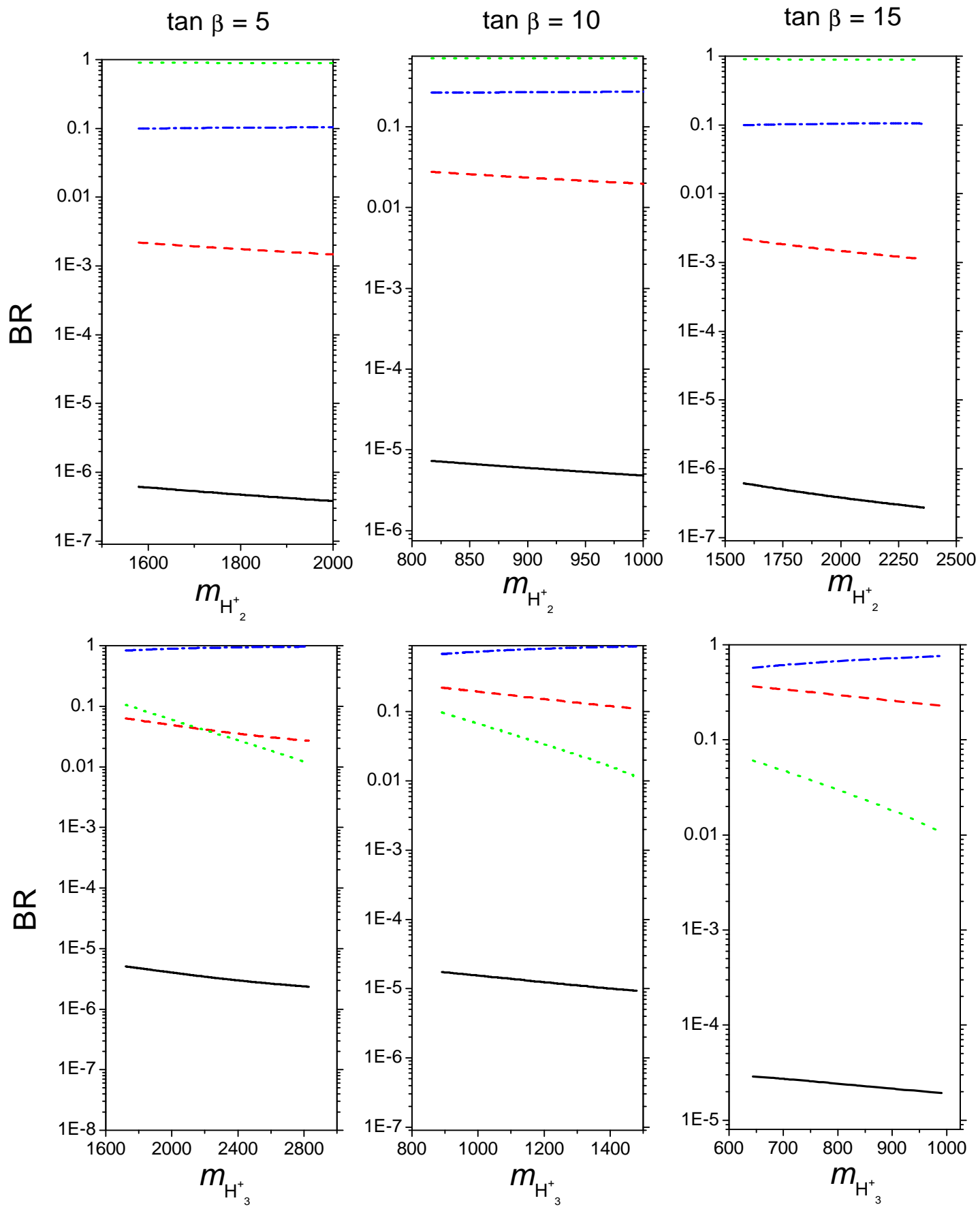
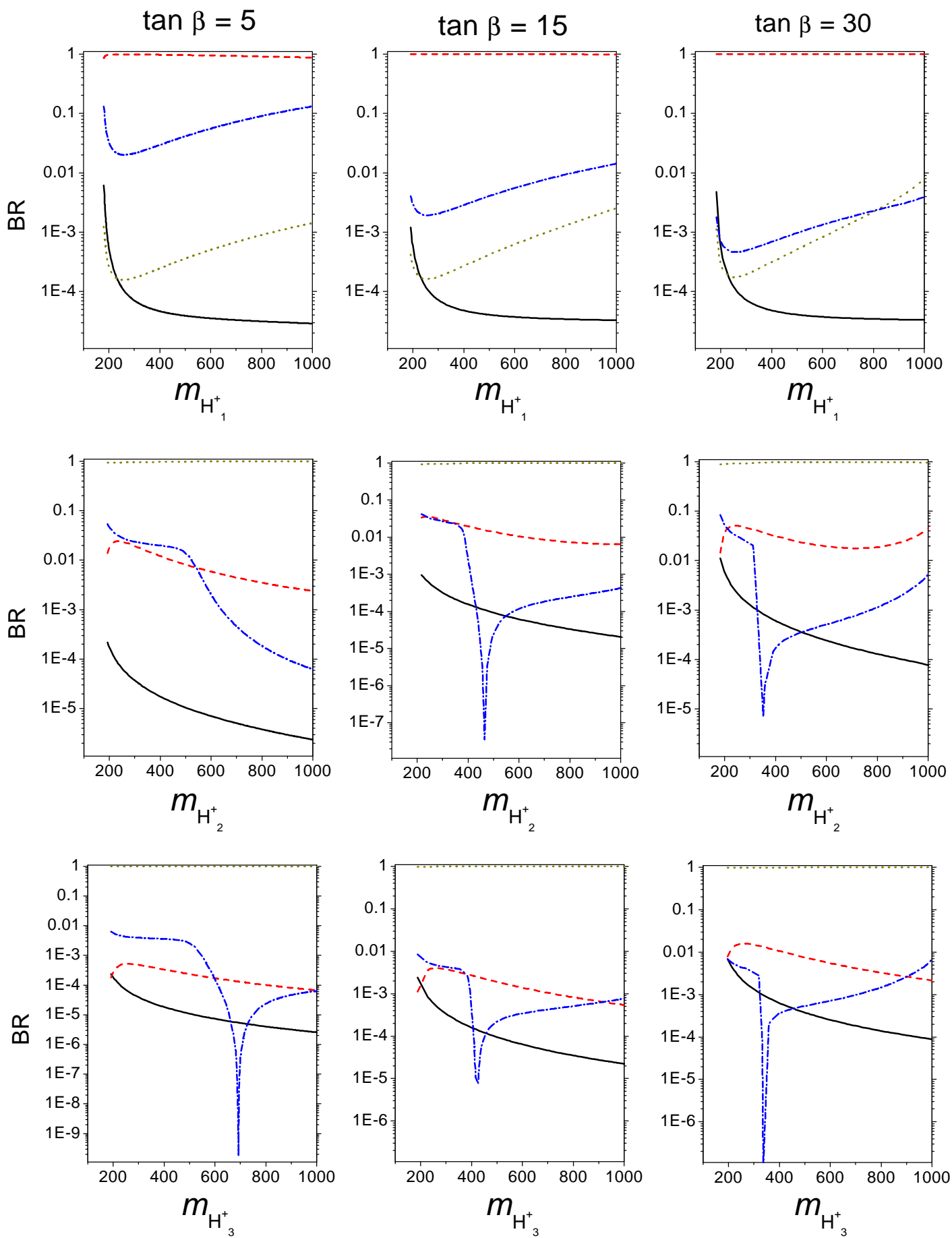
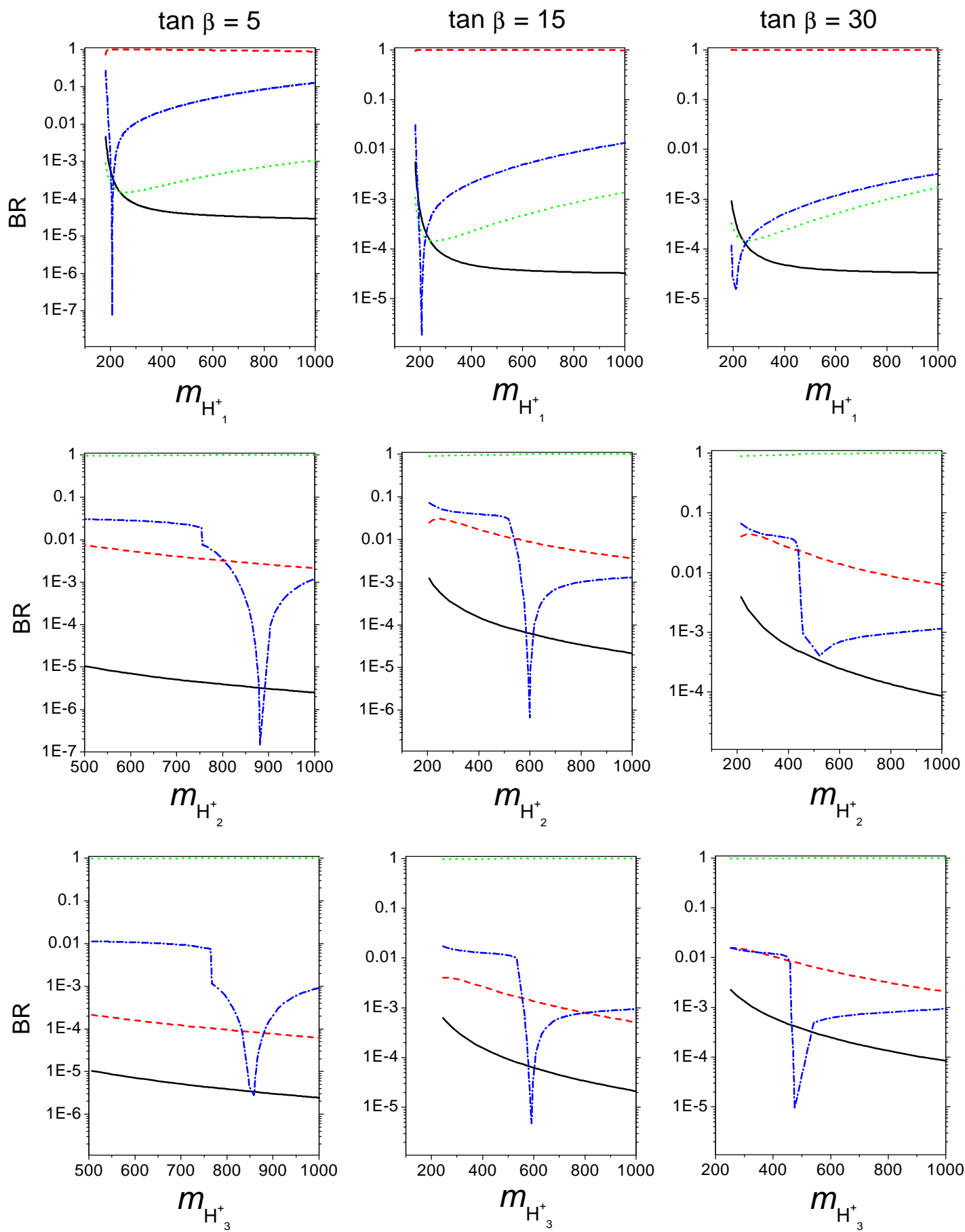
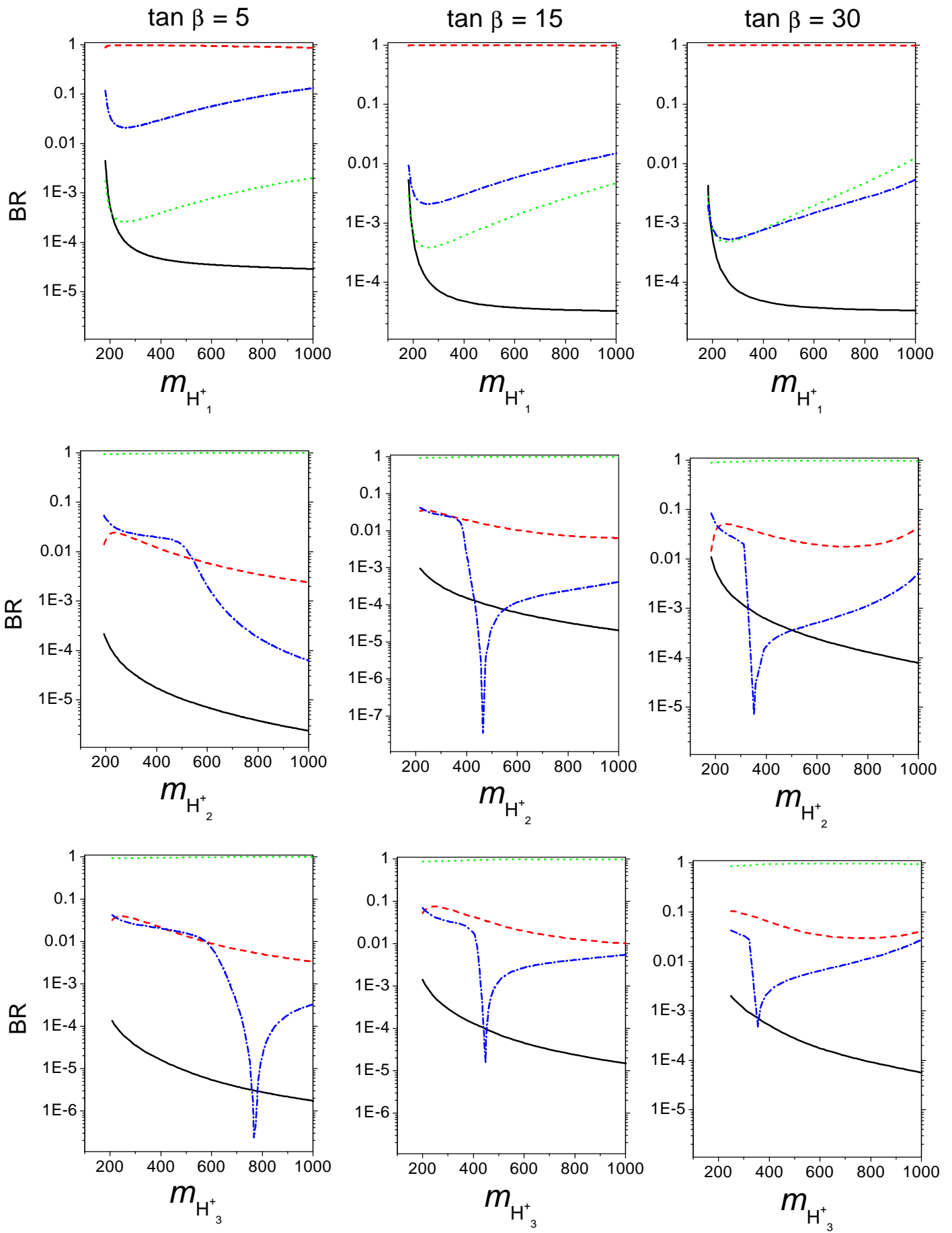
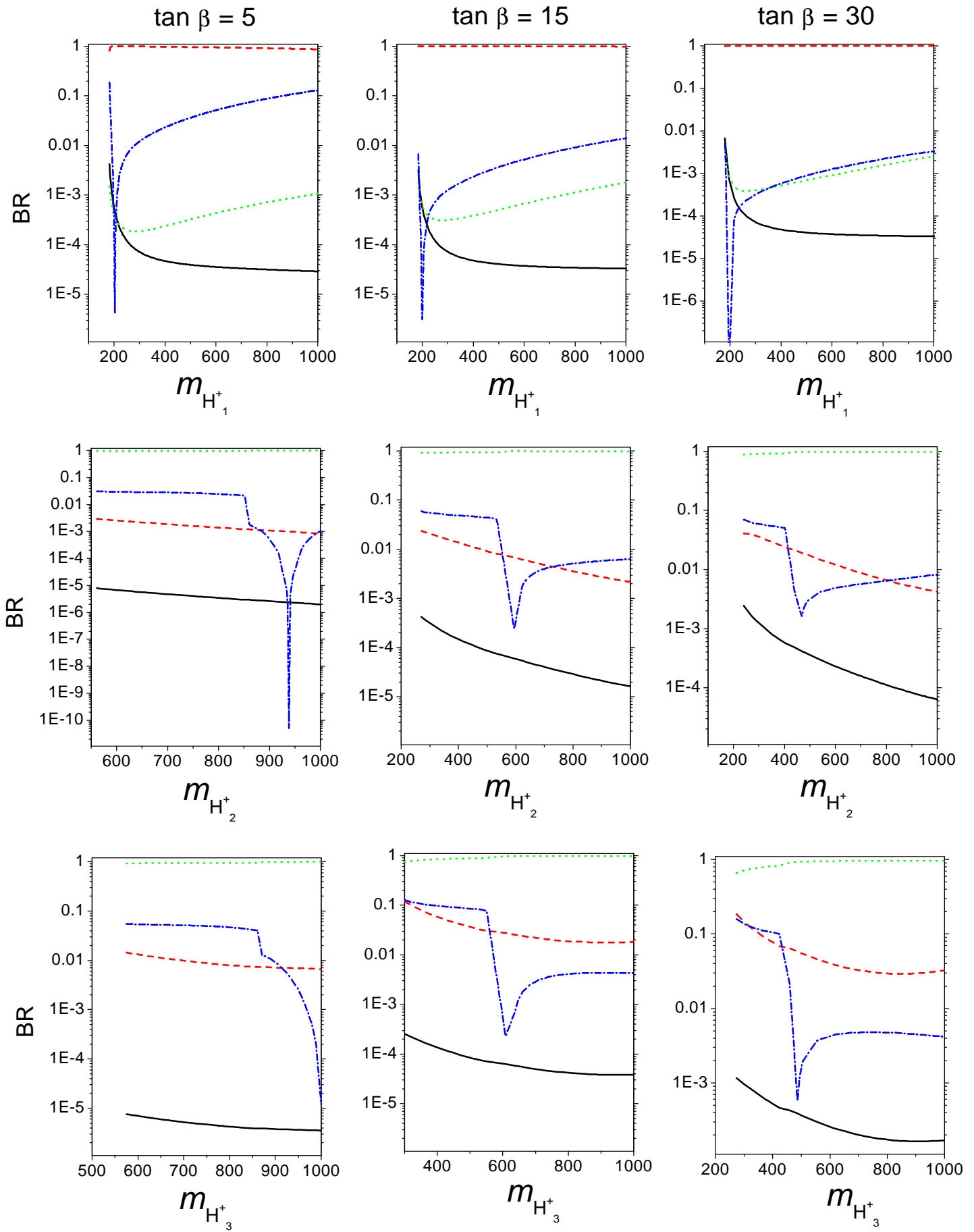
\end{document}